\documentclass[aip,jcp,amsmath,amssymb,
reprint
 ]{revtex4-1}

	\usepackage{amsmath}
	\usepackage{makeidx}
	\usepackage{amsfonts}
	\usepackage{amssymb}
	\usepackage[usenames,dvipsnames]{pstricks}
	\usepackage{subfigure}
	\usepackage{epsfig}
	\usepackage{pst-grad} 
	\usepackage{pst-plot} 
	\usepackage[colorlinks,hyperindex]{hyperref}
	\usepackage{placeins}
	\usepackage{tabto}
	\usepackage{bm}
	\hypersetup
	{
		colorlinks,%
		citecolor=black,%
		linkcolor=black,%
		urlcolor=black,%
	}
\usepackage{verbatim}

\newcommand{\Z}{\mathbb{Z}}
\newcommand{\R}{\mathbb{R}}
\newcommand{\V}[1]{\mathbf{#1}}

\newcommand{\intl}[3]{\int\limits_{#1}^{#2}\mathrm{d}#3\,}
\newcommand{\bracket}[1]{\left(#1\right)}

\newcommand{\eq}[1]{$\mathrm{Eq.}$~\eqref{#1}}
\newcommand{\eqs}[1]{$\mathrm{Eqs.}$~\eqref{#1}}
\newcommand{\figref}[1]{$\mathrm{Fig.}$~\ref{#1}}
\newcommand{\secref}[1]{$\mathrm{Sec.}$~\ref{#1}}

\newcommand{\av}[1]{\left\langle #1 \right\rangle}

\makeindex

\begin{document}

\title{Wave propagation in spatially modulated tubes}

\author{A. Ziepke}
\email{ziepke@itp.tu-berlin.de}
\author{S. Martens}
\author{H. Engel}
\affiliation{Institut f\"ur Theoretische Physik, Hardenbergstra\ss e 36, EW 7-1, Technische Universit\"at Berlin, 10623 Berlin, Germany}

\date{\today}

\begin{abstract}
We investigate wave propagation in rotationally symmetric tubes with a periodic spatial modulation of cross section.
Using an asymptotic perturbation analysis, the governing quasi two-dimensional reaction-diffusion equation can be reduced into a one-dimensional reaction-diffusion-advection equation. Assuming a weak perturbation by the advection term and using projection method, in a second step, an equation of motion for traveling waves within such tubes can be derived. 
Both methods predict properly the nonlinear dependence of the propagation velocity on the ratio of the modulation period of the geometry to the intrinsic width of the front, or pulse. As a main feature, we can observe finite intervals of propagation failure of waves induced by the tube's modulation.
In addition, using the Fick-Jacobs approach for the highly diffusive limit we show that wave velocities within tubes are governed by an effective diffusion coefficient. Furthermore, we discuss the effects of a single bottleneck on the period of pulse trains within tubes. We observe period changes by integer fractions dependent on the bottleneck width and the period of the entering pulse train.
\end{abstract}

\maketitle

\section{Introduction}

\begin{figure}[!tb]
\center
\includegraphics[width=.9\linewidth]{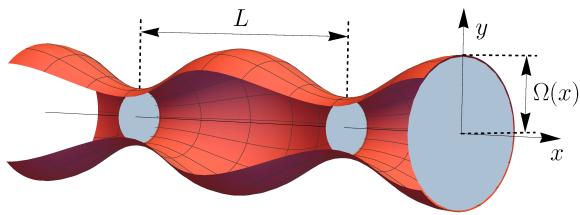}
\caption{\label{Fig:Tube} (Color online) Illustration of a rotationally symmetric tube with spatially modulated cross section $Q(x) = \pi \Omega^2(x)$ and period length $L$.}
\end{figure}

\noindent Besides the well-known Turing patterns, reaction-diffusion (RD) systems possess a rich variety of self-organized spatio-temporal wave patterns including traveling fronts, solitary excitation pulses, and periodic pulse trains in one-dimensional media. These patterns are ``building blocks'' of traveling wave patterns like target patterns, wave segments, and spiral waves \cite{Winfree1994} in two and scroll waves \cite{Keener1988} in three spatial dimensions, respectively. Traveling waves (TW) have been observed in many physical \cite{Hohenberg1993}, biological \cite{Bressloff2013}, and chemical systems \cite{Kapral1995}. Prominent examples of front propagation include catalytic oxidation of carbon monoxide on platinum single crystal surfaces \cite{Bar1992}, arrays of coupled chemical reactors \cite{Laplante1992}, and combustion reactions in condensed two-phase systems \cite{matkowsky1978propagation}. Moreover, the phenomenon of pulse propagation is associated with a large class of problems, including information processing in nervous systems \cite{koch2000}, migraine aura dynamics \cite{MigraineDahlem}, coordination of heart beat \cite{CardiacElectrophysiology}, and spatial spread of diseases \cite{DiseasesBailey}.\\
In many systems, the excitable medium supporting wave propagation exhibits a complex shape and/or is limited in size. In such cases, geometric restrictions can effect the RD processes, leading to complex wave phenomena, e.g. intracellular calcium wave patterns during fertilization of sea urchin eggs \cite{Galione1993} and in protoplasmic droplets of Physarum polycephalum \cite{Radzuweit2013}, pattern formation in the cell cortex \cite{Bois2011,shi2013interaction}, Turing patterns in microemulsion systems \cite{vanag2001pattern}, drastic lifetime enhancement of scroll waves \cite{AzhandEPL2014,Totz2015}, and atrial arrhythmia \cite{Fenton2011}. In particular, there is experimental evidence that spatial variations of the atrial wall thickness is a significant cause of scroll wave drift \cite{kharche2015} as well as anchoring \cite{yamazaki2012}; both promoting atrial fibrillation \cite{pellman2010extracellular}. Moreover, it has been reported that the dendritic shape of nerve cells strongly affects the propagation of the cellular action potential \cite{hausser_diversity_2000}.
Furthermore, the interaction of particles with porous boundaries can cause effects like adsorption and therefore influence the properties of diffusive transport. Modeling irregularities of micropores as entropic barriers, it was shown that the interplay of diffusion and boundary-induced adsorption can be described via an effective diffusion coefficient \cite{Santamaria2012}.\\
Nowadays, well-established lithography-assisted techniques enable to design the shape of catalytic domains \cite{suzuki_diffusive_2000,Baroud2003} as well as to prescribe the boundary conditions \cite{kitahata_oscillation_2009}. These provide efficient methods to study experimentally the impact of confinement on wave propagation \cite{Toth1994,agladze_electrochemical_2001,Ginn2004}, to construct chemical logical gates \cite{Steinbock1996}, and to control or optimize the local dynamics of catalytic reactions \cite{wolff2001spatiotemporal}.\\
In our recent work \cite{Martens2015}, we have provided a first systematic treatment of how propagation of traveling waves in thin three-dimensional channels with periodically varying cross-section can be reduced to a corresponding one-dimensional reaction-diffusion-advection equation (RDAE). Using projection method, we have derived an equation of motion for the position of TWs as a function of time in the presence of the boundary-induced advection term and obtained an analytical expression for the average propagation velocity. Taking the Schl\"{o}gl model for describing front dynamics, our theoretical results predict boundary-induced propagation failure being confirmed by finite element simulations of the three-dimensional RD dynamics.
Recently, Biktasheva et al. \cite{biktasheva_drift_2015} have presented a similar approach for TWs in thin layers exhibiting sharp thickness variations in which they focus on the drift of scroll waves along thickness steps, ridges, and ditches. Experiments confirming the predicted proportionality of the drift speed on the logarithm of the height variation \cite{Steinbock2015} also verify the presented analytic approach and, thus, likewise support our analytic treatment.\\
In this work, we study the propagation of TWs, in particular traveling fronts and traveling pulses, through periodically modulated tubes like the one depicted in \figref{Fig:Tube}. Therefore, we consider the two-component FitzHugh-Nagumo model as a generic model for excitable sytems which is shortly presented in \secref{Sec:Model}. In \secref{Sec:Analytics}, we apply an asymptotic perturbation analysis to derive an equation of motion for traveling waves in tubes with spatially modulated cross sections. Section \ref{Sec:Num} is dedicated to a brief description of the numerical methods being used to solve reaction-diffusion equations in confined geometries with spatially dependent Neumann boundary conditions. Following this, we discuss the numerical results as well as results of our analytic approximation for front, \secref{Sec:ResFront}, and pulse propagation in corrugated tubes, \secref{Sec:ResPulses}. Finally, we conclude the results in \secref{Sec:Conclusion}.

\section{The FitzHugh-Nagumo model}\label{Sec:Model}
\noindent In what follows, we limit our consideration to two-component reaction-diffusion systems for the concentrations $\V{u}(\V{r},t)=(u,v)^T\in\mathbb{R}^2$ whose spatial and temporal evolution is modeled by reaction-diffusion equations (RDEs)
\begin{equation}\label{Eq:RDE}
\frac{\partial\V{u}}{\partial t}(\V{r},t)=\,\mathbb{D}\Delta\V{u}(\V{r},t)+\V{R}(\V{u}(\V{r},t)).
\end{equation}
Here, $\V{r}=\bracket{x,y,z}^T\in\mathbb{R}^3$ is the position vector, $\mathbb{D}=\mathrm{diag}(D_u,D_v)$ is the diagonal matrix of constant diffusion coefficients, $\Delta$ denotes the Laplacian operator in $\R^3$, and $\V{R}\bracket{\V{u}}\in\mathbb{R}^2$ represents the nonlinear reaction kinetics. The medium filling the tubular channel, see \figref{Fig:Tube}, is assumed to be uniform, isotropic, and infinitely extended in $x$-direction.

In this work, we use the FitzHugh-Nagumo (FHN) model \cite{{FitzHugh1961},{Nagumo1962}} as a generic model for an excitable medium%
\begin{subequations}\label{Eq:FHN_RD}
\begin{align}
\partial_t u(\V{r},t) &=\,D_u\Delta u-u\bracket{u-a}\bracket{u-1}-\gamma v,\\
b\partial_t v(\V{r},t)&=\,u-v+I_i,
\end{align}
\end{subequations}
where $u=u(\V{r},t)$ and $v=v(\V{r},t)$ are the scalar concentrations of activator and inhibitor, respectively. The activator $u$ exhibits an auto-catalytic bistable reaction kinetics with local excitation threshold $a\in\left(0,1\right)$ and is coupled linearly to the inhibitor $v$. On the other hand, the dynamics of $v$ is determined by the difference $u-v$ and some applied external current $I_i$. Moreover, we assume that the activator diffuses much faster than the inhibitor, $D_u\gg D_v \equiv 0$, resulting in $\mathbb{D}=\mathrm{diag}(D_u,0)$. The remaining parameters $b$ and $\gamma$ are positive constants of the system.

\begin{figure}[!tb]
\center
\includegraphics[width=.9\linewidth]{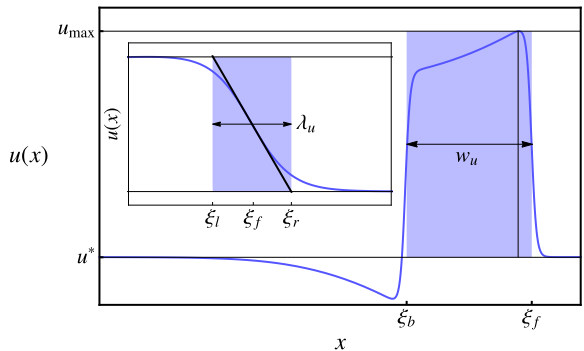}
\caption{\label{Fig:profiles} (Color online) Sketch of an activator pulse in the FHN model which can be considered as two oppositely orientated traveling fronts. The front at the side of propagation direction is located at $\xi_f$ and the coordinate $\xi_b$ shall fulfill the condition $u(\xi_b)=u(\xi_f)$ with $\xi_b\neq\xi_f$.
The pulse width $w_u= |\xi_f-\xi_b|$ is given by the absolute difference of both positions. The inset illustrates the definition of the front width $\lambda_u$ based on linearization of the pulse profile at the front position $\xi_f$. The same procedure is used to define the width of a Schl\"ogl front, identifying $l=\lambda_u$ and $x_f=\xi_f$.}
\end{figure}

First, we focus on the limiting case of a single component RD system by setting $\gamma=0$ in \eq{Eq:FHN_RD}, yielding the Schl\"ogl model \cite{Schlogl1972}
\begin{align}\label{Eq:Schloegl}
\partial_t  u(\V{r},t)=\,D_u\Delta u-u\bracket{u-a}\bracket{u-1}.
\end{align}
A linear stability analysis of the system reveals that $u^*=0$ and $u^{**}=1$ are stable spatially 
homogeneous steady states (HSS) while the local excitation threshold $u^{***}=a$ represents an unstable HSS. 
In an infinitely extended tube with non-modulated cross-section, $Q(x)=\mathrm{const}$, and a straight center line in $x$-direction, i.e., the tube is neither curved nor twisted, the Schl\"ogl model possesses a stable traveling front solution whose profile is given by
\begin{align}\label{Eq:frontprofile}
u(\V{r},t)=\,U_c(\xi ) = \left. 1\middle/ \left(1+\exp \left(\xi/\sqrt{2\,D_u}\right)\right)\right. ,
\end{align}
in the frame of reference, $\xi=x-c_0\,t$, co-moving with the free velocity $c_0$. For a Schl\"ogl front the latter is given by
\begin{align}\label{Eq:SchloeglVel}
c_0=\sqrt{\frac{D_u}{2}}(1-2a).
\end{align} 
The front solution above represents a heteroclinic connection between the two stable HSS for $\xi\rightarrow\pm\infty$, viz. $\lim_{\xi\to-\infty}u(\xi)=u^{**}$ and $\lim_{\xi\to \infty}u(\xi)=u^{*}$, and travels with $c_0$ in positive direction of the $x$-axis. The width of the traveling front
\begin{align} \label{Eq:SchloeglFW}
 l= \sqrt{32\,D_u}
\end{align}
defines the intrinsic length scale, see inset in \figref{Fig:profiles}. Noteworthy, in our scaling the front width $l$ depends solely on the diffusion constant $D_u$ and hence we can easily adjust the latter by changing the value of $D_u$ in our simulations. Further, we will refer to the position where the front solution attains the value $u(x_f)\equiv(u^{**}+u^*)/2=0.5$ as the \textit{front position} $x_f$.\\ 
Additionally, we will focus on TW solutions to \eqs{Eq:FHN_RD} propagating with constant free velocity $c_0$, viz. single solitary pulses and spatially periodic pulse trains.
For the sake of simplicity, the profile of an activator pulse $u(x)$ can be described by the combination of two oppositely orientated traveling fronts propagating in positive $x$-direction with velocity $c$. Thus, we can introduce the front position $\xi_f$ of an activator pulse in a similar way to the Schl\"ogl model, viz. we define the front position of a pulse as the location where the concentration field equals half the sum of the maximum value $u_{\mathrm{max}}$ and the HSS $u^*$ with $u^*=\lim_{x \to \pm \infty} u(x)$, $u(\xi_f)\equiv (u_{\mathrm{max}}+u^*)/2$. For a single pulse, there exist two values of $x$ for which this condition holds. The one that is further towards the direction of pulse propagation, we want to refer to as the front position $\xi_f$. The other value represents the back position $\xi_b$.
We define the pulse's front width $\lambda_u$ by linearization at $\xi_f$ and measuring the distance between the points where the linear fit function $f_\mathrm{fit}(x)$ equals the maximum value $f_\mathrm{fit}(\xi_l)=u_{\mathrm{max}}$ and the HSS $f_\mathrm{fit}(\xi_r)= u^*$, i.e., $\lambda_u=\xi_r-\xi_l$. Moreover, we define the width of the activator pulse $w_u= |\xi_f-\xi_b|$ as the absolute difference between front and back, cf. \figref{Fig:profiles}.
As the analytic solution for the unperturbed pulse profile as well as the free propagation velocity $c_0$ are unknown, these quantities have to be measured numerically for a given parameter set.


\section{Analytic Approximation}\label{Sec:Analytics}
Following our recent paper \cite{Martens2015}, we derive an equation of motion (EOM) for TWs propagating through $3$D cylindrical tubes with periodically varying cross section $Q(x)$ as depicted in Fig. \ref{Fig:Tube}. The rotationally symmetric, $L$-periodic modulation of the tube's radius is given by $\Omega(x)$; resulting in a periodically modulated cross section $Q(x)=\pi\,\Omega(x)^2$. With respect to the geometry it is convenient to use cylindrical coordinates and the RDE, \eq{Eq:RDE}, becomes
\begin{equation}\label{Eq:RDEcyl}
 \begin{split}
 \partial_t\V{u}(\rho,\phi,x,t)=&\mathbb{D}\!\left(\!\frac{1}{\rho}\partial_{\rho}(\rho\partial_{\rho})\!+\!\frac{1}{\rho^2}\partial_{\phi}^2\!+\!\partial_x^2\!\right)\!\V{u}+\V{R}(\V{u}),
\end{split}
\end{equation}
with the tube's radial and angular coordinates $\rho$ and $\phi$, respectively,
\begin{equation*}
x\in\mathbb{R},\quad \phi\in\left[0,2\pi\right),\mbox{ and}\quad \rho\in\left[0,\Omega(x)\right].
\end{equation*}
We assume impermeability of the tube walls with respect to diffusion, so that the gradient of $\V{u}$ shall obey no-flux boundary conditions (BCs), \mbox{$\bracket{\nabla \V{u}}\cdot \V{n}(x)=0,$} at the boundary $\,\rho=\Omega(x)$ with the outward pointing normal vector $\V{n}=-\Omega'(x)\V{e}_x+\V{e}_{\rho}$, yielding
\begin{subequations}\label{Eq:BCs}
\begin{align}
0=&\,-\Omega'(x)\partial_x \V{u}+\partial_\rho \V{u}, \quad \mbox{at}\,\rho=\Omega(x).
\intertext{The prime denotes the derivative with respect to $x$ and $\bm{e}_i$ represents the unit vector in the direction of $i$. 
For further simplification, we assume an angular symmetry of the initial preparation and the chosen geometry.
As a result of this restriction the diffusive material flux $\nabla \V{u}$ must be parallel with the tube's centerline at $\rho=0$,}
0=&\,\partial_\rho \V{u}(\rho,x,t), \quad \mbox{at}\,\rho=0.
\end{align}
\end{subequations}
Below, we shortly discuss the major steps in asymptotic perturbation analysis for deriving the EOM for TWs in spatially modulated tubes. The key assumption is that the modulation of the tube's radius is a small quantity compared to the period length $L$ of the modulation and hence we introduce the dimensionless parameter $\epsilon=(\Omega_{\text{max}}-\Omega_{\text{min}})/L\ll 1$ \cite{{Martens2011},{Martens2013}}. The latter characterizes the deviation of a modulated tube $\Omega(x)$ from a tube with constant diameter, i.e. $\epsilon=0$. Re-scaling all quantities in radial direction, $\rho\rightarrow\epsilon\rho$ and $\Omega(x)\rightarrow\epsilon h(x)$, yields
\begin{subequations} \label{Eq:ProblemScaled}
\begin{align}
\partial_t\V{u}=&\,\frac{\mathbb{D}}{\epsilon^2} \left(\partial_{\rho}^2+\frac{1}{\rho}\partial_{\rho}+\epsilon^2\partial_x^2\right)\!\V{u}+\V{R}(\V{u}), \label{Eq:RDENew}\\
0=&\,-\epsilon^2\,h'(x)\partial_x \V{u}+\partial_\rho \V{u}, \quad \mbox{at}\,\rho=h(x),\label{Eq:BCNew1}\\
0=&\,\partial_\rho \V{u}, \quad \mbox{at}\,\rho=0.\label{Eq:BCNew2}
\end{align}
\end{subequations}
Since \eqs{Eq:ProblemScaled} only contain terms in even orders of $\epsilon$, we expand the concentration vector $\V{u}$ in a series in even orders of $\epsilon$, 
$\V{u}(\V{r},t)=\V{u}_0(\V{r},t)+\epsilon^2\V{u}_1(\V{r},t)+\mathcal{O}(\epsilon^4)$. Substituting this ansatz into \eqs{Eq:ProblemScaled}, we obtain a hierarchic set of coupled partial differential equations. In leading order, one has to solve $\mathbb{D}\rho^{-1}\partial_{\rho}(\rho\partial_{\rho})\V{u}_0=0$ supplemented with the BCs, $0=\partial_\rho \V{u}_0$ if $\rho\in\{0,h(x)\}$, resulting in the formal solution $\V{u}_0=\V{a}_0(x,t)$. Noteworthy, the zeroth order solution is independent of the radial extend $h(x)$ and the unknown function $\V{a}_0(x,t)$ has to be determined from the second order $\mathcal{O}\bracket{\epsilon^2}$ balance, \eq{Eq:RDENew}. Integrating the latter over the re-scaled local cross section $\rho \mathrm{d}\rho\,\mathrm{d}\phi$ and taking into account the corresponding BCs, \eqs{Eq:BCNew1}-\eqref{Eq:BCNew2}, one obtains a one-dimensional reaction-diffusion-advection equation%
\begin{align} \label{Eq:RDA}
\partial_t\V{u}_0=\mathbb{D}\partial_x^2\V{u}_0+2\frac{\Omega'(x)}{\Omega(x)}\mathbb{D}\partial_x\V{u}_0+\V{R}(\V{u}_0).
\end{align} 
To sum up, the quasi two-dimensional problem with spatially dependent Neumann boundary conditions on the reactants, \eqs{Eq:RDEcyl}-\eqref{Eq:BCs}, translates into a one-dimensional RDAE with a \textit{boundary-induced advection term}, \eq{Eq:RDA}, by applying asymptotic perturbation analysis in the small parameter $\epsilon$. The advective velocity field $\V{v}=\,Q'(x)/Q(x)\V{e}_x=\,2\Omega'(x)/\Omega(x)\V{e}_x$ reflects the periodicity $L$ of the tube's modulation, $\V{v}(x+L)= \V{v}(x)$, and has zero mean, $\int_{0}^{L}\mathrm{d}x\,\V{v}(x)=\V{0}$. For systems where diffusion \cite{keener2000propagation}, advection \cite{matkowsky1978propagation}, and reaction coefficients \cite{Loeber2012} depend periodically on space and time it has been shown that the profile of a traveling front and its current velocity change periodically in time \cite{xin_front_2000} -- the traveling front solutions $u(x,t)$ are called \textit{pulsating traveling fronts} (PTFs)
\begin{align}
u\bracket{x,t+k\frac{L}{c}}=\,u\bracket{x-k\,L,t},\quad \forall k \in \Z,
\end{align}
propagating in direction of the $x$-axis with an average velocity $c$. A lot of work has been done to proof the existence and stability of these PTFs \cite{nadin2010} and to calculate the minimal speed of PTFs by a variational formula \cite{berestycki2005speed}.

Despite the fact that the boundary-induced advection term is proportional to $Q'(x)/Q(x)$ for rotationally symmetric tubes as well as thin $3$D channels with modulated rectangular cross section \cite{Martens2015}, we emphasize that, identifying the tube's diameter with the width of a planar channel, the amplitude of the advection field is two times larger for tubes, $Q(x)=\,\pi \Omega^2(x)$, compared to channels with rectangular cross section, $Q(x)=\,2\Omega(x)\,H$; here, $H$ denotes the height of the thin $3$D channel. Consequently, we expect a much stronger impact of the tube's modulation on the propagation properties of TWs.

Applying projection method \cite{Loeber2014PRL,Martens2015} to \eq{Eq:RDA}, it is feasible to derive the EOM for the position $\varphi(t)$ of TWs in response to the advection term $\mathbb{D}\bracket{\V{v}\cdot\nabla}\V{u}_0$. Assuming the latter represents a weak perturbation to a stable TW solution $\V{U}_c$ of the RDE, \eq{Eq:RDE}, one gets
\begin{align}\label{Eq:EOM}
\dot{\varphi}=c_0-\frac{2}{K_c}\int_{-\infty}^{\infty}\V{W}^{\dagger}(\xi)^{T}\mathbb{D}\frac{\Omega'(\xi+\varphi(t))}{\Omega(\xi+\varphi(t))}\V{U}_c'(\xi)\mathrm{d}\xi,
\end{align}
with the constant $K_c=\int_{-\infty}^{\infty}\mathrm{d}\xi\V{W}^{\dagger}(\xi)^T\V{U}_c'(\xi)$, initial condition $\varphi(t_0)=\varphi_0$ and a dot denoting the derivative with respect to time. Thereby, $\V{U}_c'(\xi)$ is the vector of eigenfunctions of the linearized operator $\mathcal{L}=\,\mathbb{D}\partial_\xi^2+c_0\partial_\xi +\mathcal{D}\V{R}\bracket{\V{U}_c}$ to eigenvalue zero -- the so-called \textit{Goldstone modes}. Analogous, the \textit{response functions} $\V{W}^{\dagger}(\xi)$ are the eigenvectors of the adjoint operator $\mathcal{L}^\dagger=\,\mathbb{D}\partial_\xi^2-c_0\partial_\xi +\mathcal{D}\V{R}\bracket{\V{U}_c}^T$ to the eigenvalue zero. 
Since the integrand in \eq{Eq:EOM} does not explicitly depend on time, the mean time $T_c$ the TW needs to travel one period $L$ is given by $T_c =\, \int_{0}^{L}\mathrm{d}{\varphi} [c_0- \Theta (\varphi)]^{-1}$ and thus the average propagation velocity $c$ reads
 \begin{align}
 c=&\, \frac{L}{T_c}= L \Big/ \intl{0}{L}{\varphi} \frac{1}{c_0- \Theta (\varphi)}, \label{Eq:theo_c}
\end{align}
with substitute $\Theta (\varphi) = 2\int_{-\infty}^{\infty}\mathrm{d}\xi\,\V{W}^{\dagger T} \mathbb{D} \frac{\Omega'(\xi+\varphi)}{\Omega(\xi+\varphi)} \V{U}_c'/K_c$.

\section{Numerical Approach}\label{Sec:Num}
Today, there exist many different approaches to numerically solve PDEs on irregular domains like finite element method \cite{dhatt_finite_2012}, finite difference schemes on non-uniform regular meshes with boundary interpolation \cite{davis2013numerical}, or finite differences on Cartesian grid embedded boundary method \cite{johansen1998cartesian}, to name a few. Here, we present a different approach to solve a reaction-diffusion equation, \eq{Eq:RDE}, within an irregular domain $x \in \R , \rho \in [0,\Omega(x)]$. The method is based on a coordinate transformation to map the boundaries of the periodically modulated tube onto a couple of straight lines (rectangular grid). To do so, we construct a family of curves by introducing re-scaled cylindrical coordinates $(\tilde{x} \in \R , \tilde{\rho} \in [0,1],\tilde{\phi}\in[0,2\pi))$, with
\begin{equation}\label{Eq:Trafo}
\tilde{x}=x,\quad\tilde{\rho}=\frac{\sqrt{y^2+z^2}}{\Omega(x)},\quad \mbox{and}\,\,\tilde{\phi}=\arctan\bracket{\frac{z}{y}},
\end{equation}
where any point in the tube $\V{r}=(x,y,z)^T$ is now identified by the new coordinates $q^\mu \in \lbrace \tilde{x},\tilde{\rho},\tilde{\phi}\rbrace$. Due to the coordinate transformation, we have to transform the Laplacian and the no-flux BCs. Applying Einstein notation, the Laplace-Beltrami operator in arbitrary coordinates is given by
\begin{equation}\label{Eq:LaplaceBeltrami}
\Delta=\frac{1}{\sqrt{\left| g\right|}}\frac{\partial}{\partial q^{\mu}}\left(\sqrt{\left| g\right|}g^{\mu \nu}\frac{\partial}{\partial q^{\nu}}\right).
\end{equation}
From \eqs{Eq:Trafo}, we obtain the metric tensor $g_{\mu \nu} = \partial \V{r}/\partial q^\mu \cdot \partial \V{r}/\partial q^\nu$ of the coordinate system
\begin{align}
g_{\mu\nu}=\left(
\begin{array}{c c c}
1+\tilde{\rho}^2\,\Omega'(\tilde{x})^2 & \tilde{\rho}\Omega(\tilde{x})\Omega'(\tilde{x})&0\\
 \tilde{\rho}\Omega(\tilde{x})\Omega'(\tilde{x}) & \Omega(\tilde{x})^2 &0\\ 0&0&\tilde{\rho}^2\,\Omega(\tilde{x})^2
\end{array}
\right),
\end{align}
with the determinant of the metric tensor $\left| g\right|=\Omega(\tilde{x})^4\tilde{\rho}^2$. Further, the inverse metric tensor reads
\begin{align}
g^{\mu\nu}=\frac{1}{\Omega(\tilde{x})^2}\left(
\begin{array}{c c c}
\Omega^2	&-\tilde{\rho}\Omega'&0\\
-\tilde{\rho}\Omega'&1+\Omega'^2\tilde{\rho}^2/\Omega^2 & 0\\
0&0&\Omega^2
\end{array}
\right).
\end{align}  
Presuming rotational symmetry for any solution to \eq{Eq:RDE}, the Laplacian reads
\begin{align}\label{Eq:LaplaceTrafo}
\nonumber\Delta=&\partial_{\tilde{x}}^2-\frac{2\Omega'\tilde{\rho}}{\Omega}\partial_{\tilde{x}\tilde{\rho}}+\frac{\tilde{\rho}(2\Omega'^2-\Omega''\Omega)}{\Omega^2}\partial_{\tilde{\rho}}+\frac{1}{\Omega^2\tilde{\rho}}\partial_{\tilde{x}}\\
&+\frac{1+\Omega'^2\tilde{\rho}^2}{\Omega^2}\partial_{\tilde{\rho}}^2,
\end{align}
and the no-flux BCs, \eqs{Eq:BCs}, are given by
\begin{subequations}\label{Eq:BCmap}
\begin{align}
0&=-\Omega'\Omega\partial_{\tilde{x}}\V{u}+(1+\Omega'^2)\partial_{\tilde{\rho}}\V{u},\,&\mbox{at }\tilde{\rho}=1,\\
0&=\partial_{\tilde{\rho}}\V{u},\, &\mbox{at }\tilde{\rho}=0.
\end{align}
\end{subequations}
By applying the transformation, the irregular domain inside the tube is mapped onto a non-tilted rectangular regime  \cite{Dagdug2015}. The price paid is that the Laplacian becomes a stiff elliptic operator with spatial-dependent coefficients.
The additional derivatives ($\tilde{x}$ and $\tilde{\rho}$ are skew coordinates) and the spatial dependence of the factors in the Laplacian as well as in the BCs unveil the disadvantages of the chosen coordinate transformation one has to accept in order to use a regular grid for numerical integration with finite differences.

In our numerics, we use a semi-backward Euler algorithm for integrating the RDE in the new coordinate system. In particular, we solve the matrix equation
\begin{align}\label{Eq:NumSolve}
\mathbb{M}\V{A}^{t+dt}=\V{A}^t+\mathrm{d}t\V{\Gamma}\left(A^t\right)
\end{align}
for every individual species at time $t$ with the numerical time step $\mathrm{d}t$. The vector $\V{A}^{\tau}\in\mathbb{R}^{mn}$ includes the concentration of a given chemical species at all points of the finite difference grid composed of $m$ points in $\tilde{x}$- and $n$ nodes in $\tilde{\rho}$-direction at time step $\tau$. Analogously, the vector $\V{\Gamma}(A^{\tau})\in\mathbb{R}^{mn}$ consists of the values of the reaction function $R$ at every grid point. The square matrix $\mathbb{M}=\mathbb{I}-D\mathrm{d}t\Delta_{\text{Mat.}}\in\text{Mat}(mn\times mn,\mathbb{R})$ includes the identity $\mathbb{I}$ and a matrix representation $\Delta_{\mathrm{Mat.}}$ of the Laplacian, Eq. \eqref{Eq:LaplaceTrafo}, with a compact 9-point stencil for finite differences. The item $D\in\{D_u,D_v\}$ denotes the species' diffusion coefficient. We note that $\mathbb{M}$ is a sparse matrix and the linear part of \eq{Eq:NumSolve} was solved using UMFPACK \cite{Davis2004}.

\section{Front propagation in sinusoidally modulated tubes}\label{Sec:ResFront}
Next, we study the impact of the modulation of the tube's diameter on the propagation velocity of traveling fronts. Therefore, the Schl\"ogl model, \eq{Eq:Schloegl}, supplemented with the Neumann BCs, \eqs{Eq:BCs}, is solved numerically using the method described previously in \secref{Sec:Num}. 
The data for the average front velocity $c$ are determined from a linear fit to a front position vs. time plot after subtracting transients, $c=\lim_{t\to \infty} x_f(t)/t$. In particular, we test our analytic estimate for the average front velocity $c$ of a TW using the RDAE, \eq{Eq:RDA}, with numerical results. 

For the profile of the tube's radial extend, we choose a sinusoidally modulated boundary function with period $L$
\begin{align}\label{Eq:Modulation}
\Omega(x)=0.5\left[1+\delta+(1-\delta)\sin\left(\frac{2\pi x}{L}\right)\right].
\end{align}
The maximum radius is set to $\Omega_\mathrm{max}=1$ and $\delta$ denotes the ratio of the tube's bottleneck width $2\Omega_{\mathrm{min}}$ to the maximum diameter $2\Omega_\mathrm{max}$, i.e. $\delta=\Omega_\mathrm{min}/\Omega_\mathrm{max}=\Omega_{\mathrm{min}}$. The chosen boundary profile can be seen as the first harmonic of a Fourier series of an arbitrary periodic boundary profile.

\begin{figure}[!bt]
\center
\includegraphics[width=.9\linewidth]{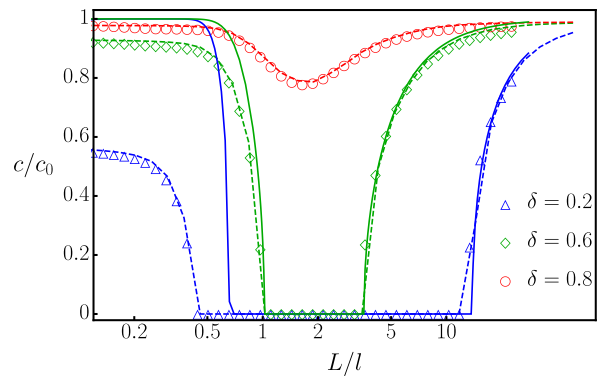}
\caption{\label{Fig:Corrugation} (Color online) Average front velocities $c$ in units of the free velocity $c_0$ versus the ratio of period length $L$ to the intrinsic front width $l$ for a sinusoidally modulated tube; see \figref{Fig:Tube}. The markers represent results from numerical simulations of the full system, \eqs{Eq:Schloegl}, \eqref{Eq:LaplaceTrafo} and \eqref{Eq:BCmap}, for different values for the aspect ratio $\delta=\Omega_{\mathrm{min}}/\Omega_{\mathrm{max}}$. Dashed lines illustrate the results for $c/c_0$ based on the RDA equation, \eq{Eq:RDA}, and solid lines correspond to the projection method, \eq{Eq:theo_c}. The remaining parameter values are set to $L=5$ and $a=0.4$.}
\end{figure}
In \figref{Fig:Corrugation}, the average front velocity $c$ in units of the unperturbed front velocity $c_0$, \eq{Eq:SchloeglVel}, as a function of the ratio of the modulation's period length $L$ to the intrinsic front width $l$ is shown. In order to adjust the ratio $L/l$, the period length is kept fixed at $L=5$ and the front width is varied by changing the diffusion constant $D_u$, see \eq{Eq:SchloeglFW}. This assures that the value for the expansion parameter $\epsilon$ stays constant for a given aspect ratio $\delta$, $\epsilon=(1-\delta)/L$, and thus allows us to verify the quality of asymptotic perturbation analysis leading to the $1$D RDAE, \eq{Eq:RDA}. 

We observe a nonlinear dependence of $c$ on the ratio $L/l$ in \figref{Fig:Corrugation}: If the front width is much smaller compared to the modulation's period, $L/l\gg 1$, the average front velocity $c$ converges to $c_0$ for any aspect ratio $\delta>0$. With decreasing ratio $L/l$, the ratio $c/c_0$ lessens until it attains its minimum value at $L \simeq l$, starts to grow again and finally saturates at a value smaller than unity. It turns out that the minimum value as well as the saturation value depend crucially on the geometric aspect ratio $\delta$. In general, we find that the velocity $c$ diminishes with shrinking ratio $\delta$ for a given ratio $L/l$. Similar to front propagation in thin corrugated channels \cite{Martens2015}, we identify a finite interval of $L/l$ values where \textit{propagation failure} (PF) occurs, i.e. the initially traveling front becomes quenched \cite{xin_front_2000} and $c$ goes to zero. One notices that the width of the PF interval grows with decreasing value of $\delta$ and it is much broader compared to our previously studied setup \cite{Martens2015} due to the two times stronger impact of the boundary-induced advection term in tubes, $\V{v}\propto 2\,D_u$. In contrast to quasi $2$D channels, PF appears even for weakly modulated tubes with large aspect ratios $\delta \leq 0.6$. Moreover, we emphasize that the upper border where the interval of PF ends, $(L/l)_\mathrm{up} \gg 1$, can be well estimated by utilizing the eikonal approach together with the stability criteria derived by Gindrod et. al \cite{Grindrod1991} (not shown explicitly); for more details see Ref. \cite{Martens2015}.

Additionally, we compare the analytical predictions based on the $1$D RDAE (dashed lines), \eq{Eq:RDA}, and the projection method (solid lines), \eq{Eq:theo_c}, with numerical results (markers) in \figref{Fig:Corrugation}. Noteworthy, the analytic results obtained by the RDAE agree excellently with the numerics for the entire range of $L/l$ values and for all aspect ratios $\delta$. In particular, it reproduces correctly the interval of PF for intermediate values of $L/l$ as well as the saturation value of $c/c_0$ for $L/l \to 0$. In comparison of the theoretical results using projection method, \eq{Eq:theo_c}, with the numerical results one notices that \eq{Eq:theo_c} yields correct results for $L \gg l$ for any aspect ratio $\delta$ and reproduces well the interval of PF for intermediate channel corrugations $\delta \simeq 0.6$, however, it fails for small ratios $L/l \ll 1$. This is in compliance with the assumptions made to derive \eq{Eq:theo_c}; namely, the boundary-induced advection term $2\,D_u\,\mathrm{max}(|Q'(x)|) \propto D_u\,(1-\delta) \ll 1$ represents a weak perturbation to a stable TW solution $\V{U}_c$. Consequently, decreasing the ratio $L/l\propto 1/\sqrt{D_u}$ while keeping the period $L$ fixed results in larger magnitude of the perturbation and eventually leads to bigger deviations between the numerics and the projection method.

\begin{figure}[!tb]
\center
\includegraphics[width=.9\linewidth]{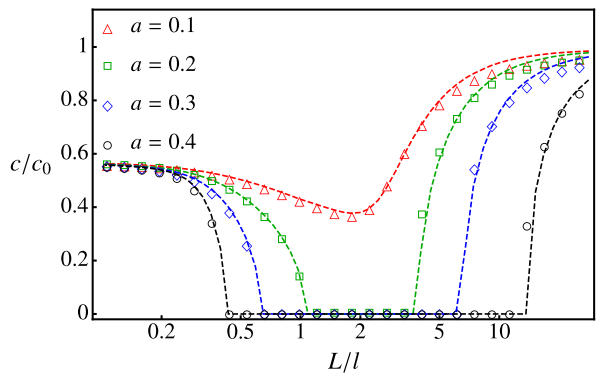}
\caption{\label{Fig:ExThresh} (Color online) Numerical results (markers) for average velocity $c$ versus $L/l$ for various excitation thresholds $a$. The analytic prediction for $c/c_0(L/l)$ based on \eq{Eq:RDA} is represented by dashed lines and yields excellent agreement. The aspect ratio and the period length of the sinusoidally modulated boundary are set to $\delta=0.2$ and $L=5$, respectively.} 
\end{figure}

In \figref{Fig:ExThresh}, we illustrate the impact of the excitation threshold parameter $a$ on the average front velocity $c$. Similar to the unperturbed front velocity $c_0$, \eq{Eq:SchloeglVel}, whose value increases for decreasing value of $a<0.5$, we observe that lowering the excitation threshold while keeping the tube parameters $L$ and $\delta$ constant facilitates the traveling front to propagate through the corrugated medium. Thus, systems with low to moderate excitability, $0.5 > a \gtrsim 0.2$, exhibit a finite interval of PF which disappears for $a\to 0$. Additionally, one observes that the normalized front velocity $c/c_0$ converges to an identical saturation value for $L/l\to 0$.

The results shown in Figs.~\ref{Fig:Corrugation} and \ref{Fig:ExThresh} indicate that this saturation value depends solely on the geometry of the tubular domain for traveling fronts whose intrinsic width, $l\propto \sqrt{D_u}$, is much larger than the period of the modulation. Is this limit, the front is extended over many periods and boundary interactions play a subordinate role. Then, the diffusion of reactants in propagation direction under spatially confined conditions is the predominant process for wave propagation and the problem can be approximated by an effective one-dimensional description introducing effective diffusion constants $\mathbb{D}_\mathrm{eff}=\mathrm{diag}(D_\mathrm{eff},0)$; yielding $\partial_t \V{u}(\V{r},t) = \mathbb{D}_\mathrm{eff} \partial_x^2 \V{u} + \V{R}(\V{u})$. Experimental \cite{Verkman2002,Keyser2014} and theoretical studies \cite{Martens2013,Burada2008} on particle transport in micro-domains with obstacles \cite{Dagdug2012a,Martens2012} and/or small openings revealed non-intuitive features like a significant suppression of particle diffusivity -- also called confined Brownian motion \cite{Cohen2006}. Numerous research activities in this topic led to the development of an approximate description of the diffusion problem -- the \textit{Fick-Jacobs approach} \cite{Zwanzig1992}. The latter predicts that the effective diffusion constant in longitudinal direction is solely determined by the variation of the cross section $Q(x)$ and can be calculated according to the Lifson-Jackson formula \cite{Lifson1962}
\begin{align}
\label{Eq:LifsonJackson}
\frac{D_{\mathrm{eff}}}{D_u}=\,\frac{1}{\langle Q(x)\rangle_{L}\langle Q^{-1}(x)\rangle_{L}},
\end{align}
where $\av{\bullet}_{L}=L^{-1}\int_{0}^{L}\bullet\,\mathrm{d}x$ denotes the average mean over one period of the modulation. For the studied tube geometry, \eq{Eq:Modulation}, the effective diffusion coefficient $D_\mathrm{eff}$ is estimated by \cite{Martens2011b}
\begin{align} \label{Eq:Deff}
D_{\mathrm{eff}}=\,\frac{16\,\delta^{3/2}\,D_u}{(3\delta^2+2\delta+3)(1+\delta)}.
\end{align}
Similar to the derivation of the RDAE for $\V{u}_0$, see \secref{Sec:Analytics}, the Fick-Jacobs approach is valid solely for weakly modulated tube geometries, i.e. $\mathrm{max}|Q'(x)| \propto \epsilon \ll 1$. 

The heuristic arguments presented above can also be confirmed by homogenization theory. As presented in Ref. \cite{Martens2016}, a rapidly, periodically changing boundary-induced advection term $Q'(x/(L/l))/Q(x/(L/l))$ in the $1$D RDAE, \eq{Eq:RDA}, can be incorporated by an effective diffusion coefficient $D_\mathrm{eff}$. The obtained expression for the latter is in complete coincidence with the result of the Lifson-Jackson formula, \eq{Eq:LifsonJackson}.

\begin{figure}[!tb]
\center
\includegraphics[width=0.9\linewidth]{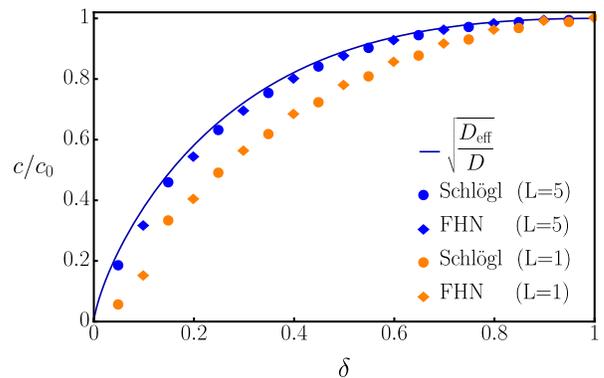}
\caption{\label{Fig:Fick} (Color online) Numerical results for average front (circles) and pulse (diamonds) velocities $c$ as a function of the aspect ratio $\delta$ of a sinusoidally modulated tube, \eq{Eq:Modulation}, with period length $L\in\{1,5\}$. The analytical prediction based on the Lifson-Jackson formula, \eq{Eq:Deff}, is represented by the solid line. The length scale ratios are set to $L/l=L/\lambda_u=0.1$ and the diffusion coefficients $D_u$ are adjusted accordingly. The remaining parameters are $a=0.4$, $b=100$, $\gamma = 0.125$, $I_i = -0.3$, and $\Omega_{\mathrm{max}}=1$.}
\end{figure}

Figure \ref{Fig:Fick} depicts the average front velocity versus the tube's aspect ratios $\delta$ for the limit $L/l \to 0$. According to the Luther's law \cite{Luther1906}, the propagation velocity depends on the square root of the effective diffusion coefficient, $c \propto \sqrt{D_\mathrm{eff}}$. One notices that the analytic estimate using $D_\mathrm{eff}$, \eq{Eq:Deff}, agrees excellently with our simulation results (markers) for weakly modulated channels ($L=5$), which confirms the heuristic explanation given above. At fixed maximum width, the front and pulse velocities increase for wider bottleneck widths, $\delta=\Omega_\mathrm{min}$, and approach the free velocities $c_0$ if the modulation disappears. 

For shorter periods, $L=1$, or stronger tube modulations, $\epsilon \in [0,1)$, one observes deviations between the numerical results and the analytic prediction, \eq{Eq:Deff}. In this limit, higher order corrections in $\epsilon$ to the effective diffusion coefficient are necessary in order to ensure a good agreement between numerics and analytics \cite{Martens2011b}.

\section{Results for traveling pulses}\label{Sec:ResPulses}
\FloatBarrier
Below, we present our results for pulse propagation within modulated tubes using the FHN model, \eqs{Eq:FHN_RD}, whereby, we set the following parameters of the model: $a=0.4, b=100, \gamma=0.125$, and $I_i=-0.3$. In contrast to the investigations of traveling fronts, the unperturbed pulse widths $w_u$, the free velocities $c_0$, and the pulses' front widths $\lambda_u$ have to be measured numerically. 
\subsection{Solitary Pulses}
\begin{figure}[!bt]
\center
\includegraphics[width = .9\linewidth]{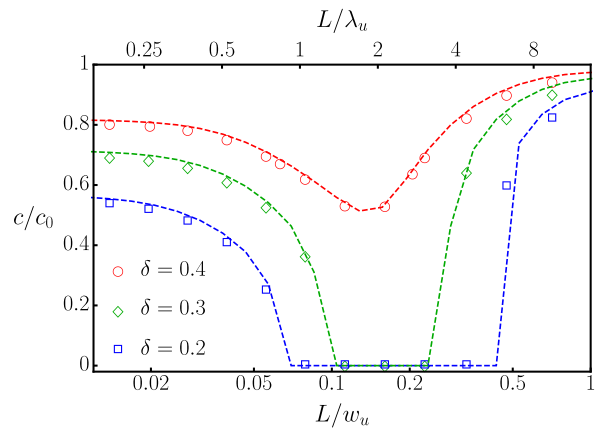}
\caption{\label{Fig:Pulses} (Color online) Average pulse velocities $c$ in units of the free velocities $c_0$ within a modulated tube for different ratios of the period length $L=5$ of the modulation to the free activator pulse width $w_u$ and the pulses' free front width $\lambda_u$. The markers represent results of numerical simulations for different values of $\delta=\Omega_{\text{max}}/\Omega_{\text{min}}$ and the dashed lines are solutions of the derived RDAE, \eq{Eq:RDA}. The remaining parameters are $a=0.4, b=100,\gamma=0.125, I_i=-0.3, \mbox{and } D_v =0$.}
\end{figure}

First, we consider a solitary pulse traveling through sinusoidally modulated tubes, \eq{Eq:Modulation}. In complete analogy to the case of front propagation, we regard the average pulse velocities $c$ in units of the free velocities $c_0$ for different activator pulse widths $w_u$ being adjusted via the corresponding diffusion coefficient $D_u$, in \figref{Fig:Pulses}.

At first glance, our numerical results show a very similar behavior compared to the results for traveling fronts, see Fig. \ref{Fig:Corrugation}. In the limit $L/w_u\to\infty$, the velocities inside the tube approach the free velocity $c_0$ regardless of the tube's corrugation. With increasing pulse width, the ratio $c/c_0$ decreases and in cases of rather strongly corrugated tubes, i.e. $\delta \leq 0.3$, again, a finite interval of PF appears.
If the activator diffusion is further increased, pulse velocities become larger again and the ratios $c/c_0$ converge to values below unity in the limit $L/w_u\to 0$. One observes a comparable behavior for pulse velocities if one varies the modulation's period length at a fixed value of the diffusion coefficient $D_u$ (not shown here).
Interestingly, in comparison with front solutions, one needs stronger modulations, $\delta < 0.4$, to observe PF for pulses, for the chosen set of parameters. One should note that the interval of PF is located at ratios of $L/w_u \sim 0.1$. But if one identifies the pulse's front width $\lambda_u$ as the relevant length scale, PF occurs at ratios comparable with those of traveling fronts, namely at $L/\lambda_u\sim L/l\sim 1$, see scale at top of Fig. \ref{Fig:Pulses}.
In addition, we emphasize, that the results of the RDAE agree excellently with those of the full simulation.\\

\subsection{Periodic Pulse Trains}
\begin{figure}[!bt]
\center
\includegraphics[width=\linewidth]{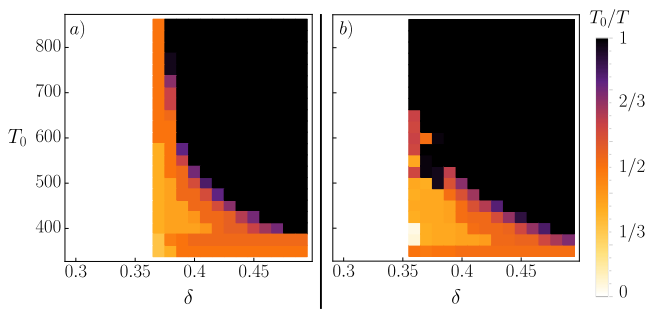}
\caption{\label{Fig:PulseTrains} (Color online) Ratio of the initial period $T_0$ of a pulse train to its period $T$ after passing a single bottleneck with diameter $2\delta$ in an angular symmetric tube with width $2$. Left panel $a)$: Results from the full numerical simulations based on \eqs{Eq:Schloegl}, \eqref{Eq:LaplaceTrafo} and \eqref{Eq:BCmap}. Right panel $b)$: Results from the numerical solution of the RDAE, \eq{Eq:RDA}. The diffusion coefficients are set to $D_u=1, D_v=0$.}
\end{figure}

In this subsection we investigate the behavior of FHN-pulse trains within angular symmetric tubes with a single sinusoidal bottleneck. In between two straight sections, the tube's geometry is chosen to be modulated according to \eq{Eq:Modulation} for one period $L=5$ which results in a single bottleneck with width $2\delta$. Pulses are initiated at one side of the tube by setting the activator values within an interval of the free pulse width $w_u$ to $1$ for one time-step $\mathrm{d}t$. No-flux BCs in $x$-direction and a fixed activator value, equal to the rest state, at the very left of the tube will cause the initiation of pulse propagation. Then, after each integer multiple of the time difference $T_0$, a pulse is initiated in the same manner.
\figref{Fig:PulseTrains} depicts the dependence of the period $T$, defined as the time difference between pulses exiting the bottleneck, on the period $T_0$ of the entering pulse train and the corrugation parameter $\delta$.

One observes a very good agreement of the results of the full numerical simulations, \figref{Fig:PulseTrains}a), and the solution of the RDAE, \figref{Fig:PulseTrains}b).
Both show that for small bottleneck widths, e.g. $\delta\leq0.36$ the ratio $T_0/T$ of the periods is zero for all initial time differences $T_0\in\left[350,850\right]$. This is related to a situation in which all pulses get pinned at the bottleneck and finally disappear. 
For periods $T_0\lesssim 375$ the pulse trains become unstable and pulses vanish even within the plain section of the tube. This fact finds verification in the numerical results, as even for broad bottlenecks the ratio $T_0/T$ does not reach unity if $T_0$ is too small. For instance, at periods $T_0\approx 375$ every second pulse vanishes and one observes a doubling of periods even before the pulses approach the bottleneck.
For large periods of the entering pulse train, $T_0>675$, the pulses can be approximately regarded as non-interacting single pulses. If the corrugation is weak enough ($\delta \gtrsim 0.36$) a single pulse is able to pass the bottleneck and so are all others of the pulse train, and hence $T=T_0$. For decreasing values of the entering period, the interaction of the pulses within the pulse train can no longer be neglected. One observes parameter regions with ratios $T_0/T\in\{1/3,1/2\}$. Also there are minor regimes in which different ratios such as $T_0/T=2/3$ occur.

As it can be seen in the RDAE, \eq{Eq:RDA}, the adoption of the pulse shape to the boundary's curvature, which is also related to the critical nucleation size after narrow gaps \cite{agladze_electrochemical_2001}, causes the pulses' front velocities to decrease in the broadening section of the tube and the pulses' back becomes decelerated while entering the bottleneck. When the next pulse approaches the region behind the previous one, it is only able to excite the medium after a given refractory time.
If, due to an incomplete recovery in the pulse's refractory tail, the region is still non-excitable after the previous pulse passed the section, the following pulse vanishes and, hence, a period change occurs. Similar effects were observed in experiments and described by Toth et al. \cite{Toth1994} who investigated the propagation of waves in the Belousov-Zhabotinsky reaction through narrow tubes connecting two reservoirs.

In addition, we want to stress that comparing the results for pulse trains with those for solitary pulses one realizes a difference in occurrence of PF. 
For $D_u=1$, PF of a single pulse occurs if $\delta<\delta_{\text{max}}\in\left(0.2,0.3\right)$, but for the same parameters, no pulses in a pulse train can pass the single bottleneck if $\delta<\delta_{\text{max}}\approx 0.36$. This condition even holds for large periods $T_0\geq 850$ and so it must be an effect of the chosen geometries.
As mentioned above, the pulses' velocity is increased in the narrowing part of the modulation and decreased if the tube expands again. If the tube exhibits an ongoing periodic modulation, the pulse senses partly an increase and a decrease of its velocity. Otherwise, if the modulation ends after a section of expanding, the pulse mainly interacts with the part of the geometry that causes a decrease of velocity. In consequence, for waves it is harder to overcome the end of a modulation, where the tube's radial extend is maximal, instead of an ongoing equivalent periodic variation of the cross section.

\section{Conclusion}\label{Sec:Conclusion}
In this work, we have investigated propagation of reaction-diffusion waves through rotationally symmetric tubes with no-flux boundaries and modulated cross section.
First, we have presented a systematic treatment of reaction-diffusion equations within such tubes by performing a multiple scale analysis to reduce the effectively two-dimensional reaction-diffusion equation to a one-dimensional reaction-diffusion-advection equation. \\
In a second step, we have obtained an analytic expression for wave velocities within the aforementioned geometries, using projection method. Moreover, we have presented a handy approach for numerical simulation of the reaction-diffusion equation within irregular geometries based on a coordinate transformation onto a rectangular integration regime.

Exemplarily using the FitzHugh-Nagumo model, we have studied the propagation of traveling front and pulse solutions in a sinusoidally modulated tube. In particular, intervals of propagation failure are found for sufficiently strong tube modulations. These finite intervals are located in parameter regions where the modulation's period length $L$ and the front width $l$, respectively the activator pulse's front width $\lambda_u$, are of the same order of magnitude. 
In the limit of narrow front widths, $L/l\rightarrow \infty$ or $L/\lambda_u\rightarrow\infty$, the wave velocities approach the free propagation velocities for any finite bottleneck width, whereas in the limit of large front widths, $L/l\rightarrow 0$ or $L/\lambda_u\rightarrow 0$, the wave velocities saturate at values below the free velocity. These velocities depend crucially on the geometry, and the boundaries' influence can be incorporated in a one-dimensional reaction-diffusion equation via an effective diffusion coefficient which can be calculated using the Lifson-Jackson formula \cite{Lifson1962}. 
For our considerations we have chosen to tune the chemical length scale instead of the modulation's period to ensure the smallness of the geometric expansion parameter $\epsilon$, see Sec. \ref{Sec:Analytics}. Nevertheless, we emphasize that one obtains similar results by adjusting the geometric length scale instead; The latter might be easier to access in most experimental realizations. 
Further, we want to highlight the very good agreement of the solutions of the derived reaction-diffusion-advection equation and the numerical simulations of the full system regarding the wave velocities within the modulated tubes. Additionally, with a few limitations, in case of front propagation, the analytically obtained approximate solution of the reaction-diffusion-advection equation predicts well the front velocities. 

Regarding periodic pulse trains, we have studied the influence of a single sinusoidal bottleneck on the pulse train's period. Dependent on the bottleneck width and the period of the entering pulse trains, one observes different ratios of the period of exiting pulse trains to the period of entering pulse trains. Due to changes in the velocity of pulse propagation at the bottleneck, subsequent pulses can reach the medium ahead before it returned to the excitable state after the previous pulse went by. This causes the pulse to vanish and therefore the pulse train's period will increase by integer ratios of the entering period. Noteworthy, the results of the full numerical simulations and the solution of the reaction-diffusion-advection equation exhibit a very good agreement.
\acknowledgments
\noindent We acknowledge financial support from the German Science Foundation DFG through the SFB 910 ``Control of Self-Organizing Nonlinear Systems''.

\end{document}